\documentclass[prl,nofootinbib,twocolumn]{revtex4}

\usepackage{amsmath}
\usepackage[bookmarks=false]{hyperref}
\usepackage{graphicx}
\usepackage{epstopdf}

\newcommand{\nn}{\nonumber}

\begin{document}

%%%%%%%%%%%%%%%%%%%%%%%%%%%%%%%%%%%%%%%%%%%%%%%%%%%%%%%%%%%%%%%%%%%%%%%%%%%%%%%%
% Title page
%%%%%%%%%%%%%%%%%%%%%%%%%%%%%%%%%%%%%%%%%%%%%%%%%%%%%%%%%%%%%%%%%%%%%%%%%%%%%%%%

\title{Comment on new physics contributions to $\Gamma_{12}^s$}

\author{Christian W. Bauer}

\author{Nicholas Daniel Dunn}

\affiliation{Ernest Orlando Lawrence Berkeley National Laboratory,
University of California, Berkeley, CA 94720
\vspace{2ex}}

\begin{abstract}
A recent measurement by the D0 collaboration finds a like-sign di-muon charge asymmetry in the $B$ system that is roughly 3$\sigma$ larger than the value predicated by the Standard Model. This suggests new physics contributing to $B-\overline{B}$ mixing. For the current central value of the CP asymmetry, the required size of $\Gamma_{12}^s$ is larger than Standard Model estimates of this quantity. In this paper, we will explore the constraints on new physics contributions to $\Gamma_{12}^s$. We show that there are two dimension six operators of Standard Model fields in the electroweak Hamiltonian whose coefficients are not constrained enough to rule out possible contributions from new physics. We argue that a more precise measurement of $\tau(B_s)/\tau(B_d)$, which is possible with currently available data, could either support or strongly constrain the existence of new physics in $\Gamma_{12}^s$.

\end{abstract}

\maketitle

%%%%%%%%%%%%%%%%%%%%%%%%%%%%%%%%%%%%%%%%%%%%%%%%%%%%%%%%%%%%%%%%%%%%%%%%%%%%%%%%
Both the D0 and the CDF collaborations have measured the like-sign di-muon charge asymmetry in the $B$ system
\begin{equation}
A_{\rm sl}^b \equiv \frac{N_b^{++}-N_b^{--}}{N_b^{++}+N_b^{--}}\,.
\end{equation}
Using 1.6 fb$^{-1}$ of data CDF obtained~\cite{CDF9015} $A_{\rm sl}^b = (8.0 \pm 9.0 \pm 6.8)\times 10^{-3}$, while the D0 collaboration recently reported~\cite{Abazov:2010hv} a result of $A_{\rm sl}^b = (-9.57 \pm 2.51 \pm 1.46)\times 10^{-3}$ with 6.1 fb${^-1}$ of data. Combining these results, one finds
\begin{equation}
A_{\rm sl}^b \equiv \frac{N_b^{++}-N_b^{--}}{N_b^{++}+N_b^{--}} = -(8.5 \pm 2.8) \times 10^{-3}\,.
\end{equation}
This value is about 3$\sigma$ away from the Standard Model (SM) prediction of $A_{\rm sl}^b = -0.2 \times 10^{-3}$. 

Since these measurements are blind as to which flavor of $B$ meson produced the two muons, $A^b_{\rm sl}$ receives contributions from the semileptonic CP asymmetries of both $B_s$ and $B_d$ mesons, which we will call $a_{\rm sl}^s$ and $a_{\rm sl}^d$, respectively. The relation between $A_{\rm sl}^b$ and the  $a_{\rm sl}^q$ is given by~\cite{Abazov:2010hv}
\begin{equation}
A_{\rm sl}^b = (0.506 \pm 0.043) a_{\rm sl}^d + (0.494 \pm 0.043) a_{\rm sl}^s\,.
\end{equation} 
D0~\cite{Abazov:2009wg} has also measured the semileptonic CP asymmetry $a_{\rm sl}^s$ directly, albeit with large uncertainties
\begin{equation}
a_{\rm sl}^s = \left(-1.7 \pm 9.1 \right) \times 10^{-3}\,.
\end{equation}
One can convert the di-muon charge asymmetry into a measurement of the semileptonic asymmetry of the $B_s$ system using input from the $B_d$ system. If one assumes no new physics contribution to $B_d$ mixing, one finds (combining with the explicit measurements)
\begin{equation}
\left(a_{\rm sl}^s\right)_{\mbox{\tiny ${\rm SM}\,a_{\rm sl}^d$}} = -(12.2 \pm 4.9) \times 10^{-3}\,,
\end{equation}
while using the measurement~\cite{Barberio:2008fa}
 $a_{\rm sl}^d = (-4.7 \pm 4.6) \times 10^{-3}$, one finds
\begin{equation}
\left(a_{\rm sl}^s\right)_{\mbox{\tiny $a_{\rm sl}^d$ meas}} = -(9.2 \pm 4.9) \times 10^{-3}\,.
\label{eq:aslsmeas}
\end{equation}
While it is probably too early to tell if this discrepancy is due to physics beyond the Standard Model or fluctuations in the data, it is certainly interesting to understand what the implications of this measurement are for new physics (NP).

There are two amplitudes each that characterize mixing in the $B_q$ systems ($q=s,d$): the off-diagonal element of the mass matrix $M_{12}^q$ and the off-diagonal element of the decay matrix $\Gamma_{12}^q$. Only the relative phase $\phi^q$ between these two amplitudes is observable, such that one can choose the three real parameters $|M_{12}^q|$, $|\Gamma_{12}^q|$ and $\phi^q$ to describe the physics. In terms of these parameters, the semileptonic asymmetry is given by
\begin{equation}
a_{\rm sl}^q = \frac{\left|\Gamma_{12}^q\right|}{\left| M_{12}^q\right|} \sin \phi^q\,.
\end{equation}

What are the values of the three parameters $|M_{12}^q|$, $|\Gamma_{12}^q|$ and $\phi^q$ in the Standard Model? This question is not easy to answer, since both $|M_{12}^q|$ and $|\Gamma_{12}^q|$ depend on non-perturbative physics, which is notoriously difficult to determine. However, much effort has been devoted to this problem, including lattice calculations to determine required bag parameters. We use the calculations from~\cite{Lenz:2006hd}, supplemented with updated values for the decay constants and the bag parameters, obtained from~\cite{latticeaverages}. Adding the uncertainties quoted in these to references in quadrature, we find
\begin{eqnarray}
|M_{12}^s|^{\rm SM} &=& (9.8 \pm 1.1) {\rm ps}^{-1}\nn\\
|\Gamma_{12}^s|^{\rm SM} &=& (0.049 \pm 0.012) {\rm ps}^{-1}\nn\\
\phi^s &=& (0.04 \pm 0.01)\,.
\label{eq:SMvalues}
\end{eqnarray}
It should be noted that the calculation of $\Gamma_{12}^q$ relies on an operator product expansion, even though the energy released is only $m_b-2m_c \sim 2 \thinspace{\rm GeV}$. Thus one might be worried of the convergence of the expansion performed~\cite{Grinstein:2001zq}.

These parameters can also be related to other physical observables. In particular, they determine the mass and width difference between $B_s$ and $\overline B_s$ mesons, as well as the time-dependent CP asymmetry $S_{\psi\phi}$. The relations are
\begin{eqnarray}
\Delta M_s &=& 2 |M_{12}^s| \nn\\
\Delta\Gamma_s &=& 2 | \Gamma_{12}^s| \cos \phi^s\nn\\
S_{\psi\phi} &=& -\sin \phi^s\,,
\end{eqnarray}
where we have assumed $\left|\Gamma_{12}^s\right|\ll\left|M_{12}^s\right|$ and $\text{arg}[-V_{ts}V^*_{tb}/V_{cs}V^*_{cb}]\approx 0$. In terms of these three observables one finds~\cite{Grossman:2009mn}
\begin{equation}
a_{\rm sl}^s = -\frac{\Delta \Gamma_s}{\Delta M_s} \frac{S_{\psi\phi}}{\sqrt{1-S_{\psi\phi}^2}}\,.
\end{equation}

The measured values for these three observables are~\cite{CDFD0combination,HFAG}\footnote{Note that the Standard Model predicts $S_{\psi\phi}$ to be very close to zero, giving another hint at physics beyond the Standard Model in the $B_s$ system.}\footnote{A recent measurement by the Belle collaboration~\cite{Esen:2010jq} finds a value for $\Delta\Gamma_s$ that is consistent with the Standard Model values of $|\Gamma^s_{12}|$ and $\phi^s$.}
\begin{eqnarray}
\Delta M_s &=& (17.78 \pm 0.12) {\rm ps}^{-1}\nn\\
\Delta \Gamma_s &=& \left(0.154 ^{+0.054}_{-0.070}\right){\rm ps}^{-1}\nn\\
S_{\psi\phi} &=& 0.69 ^{+0.16}_{-0.23} \,.
\end{eqnarray}
Using these inputs, together with the measured value of $a_{\rm sl}^s$ given in Eq.~(\ref{eq:aslsmeas}), we can extract the three theoretical parameters. We find a good fit, indicating that the measurements are compatible with one another, with result \begin{eqnarray}
|M_{12}^s| &=& (8.889 \pm 0.060) {\rm ps}^{-1}\nn\\
|\Gamma_{12}^s| &=& (0.112 \pm 0.040) {\rm ps}^{-1}\nn\\
\phi^s &=& -0.79 \pm 0.24\,.
\end{eqnarray}
From this one can see that the data prefers $|M_{12}^s|$ to be close to the SM value, while both $|\Gamma_{12}^s|$ and $\phi^s$ differ from the values given in Eq.~(\ref{eq:SMvalues}), by about 1.5$\sigma$ and 3$\sigma$ respectively. 
This is in agreement with the result of~\cite{Ligeti:2010ia}, which also found that a good fit to the data requires a non-zero phase as well as a value of $|\Gamma_{12}^s|$ higher than what is predicted in~\cite{Lenz:2006hd}. This is also compatible with the observation made in~\cite{Dobrescu:2010rh}, which found that new physics that only adds a relative phase $\phi^s$ is unable to explain the central value of the semileptonic CP asymmetry.
If we were to assume no new physics in the $B_d$ system, we would find the same value for $|M_{12}^s|$, but 
$|\Gamma_{12}^s| = (0.131 \pm 0.41)  {\rm ps}^{-1}$ and $\phi^s = -0.88 \pm 0.24$. 

Given this result, one might naturally be inclined to add new physics to $\Gamma_{12}^s$~\cite{Dighe:2010nj}.\footnote{For previous attempts to explain the CP asymmetry by new physics contributions to $M_{12}^s$, see~\cite{Dobrescu:2010rh,Randall:1998te,Buras:2010mh,Jung:2010ik}.}
In the remainder of this paper we will study the constraints on NP contributions to $\Gamma_{12}^s$ from data on the decays of $B$ mesons. The constraints we derive are in general not sensitive to ${\cal O}(1)$ factors neglected in our calculations. However, it is possible that large numeric factors could relax or avoid some constraints.
 
Any operator of the form $\bar b s R$, with $R$ being any flavor neutral set of fields with total mass below $m_{B_s}$ can contribute to $\Gamma_{12}^s$. In order to conserve energy and momentum, $R$ needs to contain at least two fields. We first consider operators which only contain light fields present in the Standard Model, but comment on the possibility of introducing new light fields towards the end of the paper. The lowest dimensional operators possible have dimension six
\begin{equation}
O_{\rm NP}^s = \bar b  s \, \bar \psi \psi\,,
\label{eq:newop}
 \end{equation}
where $\psi$ denotes any light Standard Model fermion. It is also possible to add a pair of operators, $\bar bs\bar\psi_i\psi_j$ and $\bar bs\bar\psi_j\psi_i$, such that the combination is flavor neutral. A list of the possible operators is shown in Table~\ref{tab:operators}. 
The physics of $B$ decays is described by the electroweak Hamiltonian, which is conventionally written in the form 
\begin{equation}
H \sim 4 \frac{G_F}{\sqrt{2}} \sum_i  C_i \,O_i\,.
\end{equation}
Characterizing the scale of new physics by $\Lambda_{\rm NP}$, we write the coefficients of the new operators as
\begin{eqnarray}
C_{\rm NP}^{s} &\sim& g_{\rm NP}^2 m_W^2/\Lambda_{\rm NP}^2\,.
\end{eqnarray}
\begin{table}[here]
\begin{tabular}{|c|c|c|c|}
\hline
\multicolumn{4}{|c|}{Allowed operators}\\
\hline
\multicolumn{2}{|c|}{$B_s$} & \multicolumn{2}{|c|}{$B_d$}\\
\hline
$O_{\rm NP}^{s}$ & Constr $\Gamma$ & $O_{\rm NP}^{d}$ & Constr $\Gamma$  \\
\hline
$\bar bs\bar uu$& $K^+ \pi^-$, $K^+ \pi^0$ & 
$\bar bd \bar uu$ & $\pi^+ \pi^-$, $\pi^+ \pi^0$\\
$\bar bs\bar dd$ & $K^0 \pi^+$, $K^+ \pi^0$ & 
$\bar bd \bar dd$ & $\pi^+ \pi^0$ \\
$\bar bs\bar cc$ &  & 
$\bar bd \bar cc$ & $X_d\gamma$\\
$\bar bs\bar ss$ & $\phi K^0$ & 
$\bar bd \bar ss$ & $\bar K^0 K^+$, $K^0 \bar K^0$\\
$\bar bs\bar ee$ & $K^{(*)} e^+ e^-$ & 
$\bar bd \bar ee$ & $(\pi, \rho) e^+ e^-$ \\
$\bar bs\bar \mu\mu$ & $K^{(*)} \mu^+ \mu^-$ & 
$\bar bd \bar \mu\mu$ & $(\pi, \rho) \mu^+ \mu^-$\\
$\bar bs\bar \tau\tau$ & &  
$\bar bd \bar \tau\tau$ & $\tau^+ \tau^-$ \\
$\bar bs\bar \nu \nu$ &  $K^{(*)} \bar \nu \nu$ &
$\bar bd\bar \nu\nu$ & $(\pi,\rho) \bar \nu \nu$ \\
\cline{1-2}
$\bar bs\bar sd$ & $\bar K^0 K^0$, $K^+ \bar K^0$ & 
$\bar bd \bar sd$ & $\bar K^0 \pi^+$ \\
$\bar bs\bar ds$ & $\bar K^0 \bar K^0$, $K^+\bar K^0$& 
$\bar bd \bar ds$  & $K^0 \pi^+$  \\
\cline{1-2}
$\bar bs\bar cu$ & $D_s^+\pi^-$, $K^0 D^0$& 
$\bar bd \bar cu$ & $D^+ \pi^-$\\
$\bar bs\bar uc$ & $D^-K^+$, $\bar D^0K^+$ & 
$\bar bd \bar uc$  &  \\
\hline
\end{tabular}
\caption{Possible operators of the form $\bar b q \bar \psi \psi$, with $\psi$ being an SM fermion. In the second column we show some decays that can be used to constrain each operator. The next two columns show the same for operators in the $B_d$ system, which are required to keep the $B_d$ lifetime in agreement with the $B_s$ lifetime.}\label{tab:operators}
\end{table}

The contribution of an operator $O_{\rm NP}^{s}$ to $\Gamma_{12}^s$ can be evaluated by performing an OPE. Comparing the result with the dominant contribution in the SM, arising from the operator $\bar b s \bar c c$ with Wilson coefficient $C \sim V_{cb}$, we find
\begin{equation}
\frac{\left|\Gamma_{12}^{\rm NP}\right|}{\left|\Gamma_{12}^{\rm SM}\right|} \sim \left(\frac{C_{\rm NP}^{s}}{\left|V_{\rm cb}\right|}\right)^2\,,
\end{equation}
where we have neglected phase space factors. A new $\bar bs\bar cc$ operator that can interfere with the SM operator is an exception, which we discuss in more detail below. In order for the contribution of new physics $\Gamma_{12}^{\rm NP}$ to compete with the Standard Model contributions, the Wilson coefficient of this new operator needs to satisfy
\begin{equation}
C_{\rm NP}^{s} \sim \lambda^2\,,
\label{eq:CGamma}
\end{equation}
where $\lambda$ is the Cabibbo angle $\lambda \sim 0.2$. This is satisfied if $\Lambda_{\rm NP} \lesssim g_{\rm NP}\, m_W / \lambda $. Note that we have neglected numerical factors that arise from the contractions over Dirac and color indices. While these can be substantial for certain operators, our conclusions are in general not affected by these factors. 

\begin{figure}[here]
\centering
\includegraphics[width=0.9\columnwidth]{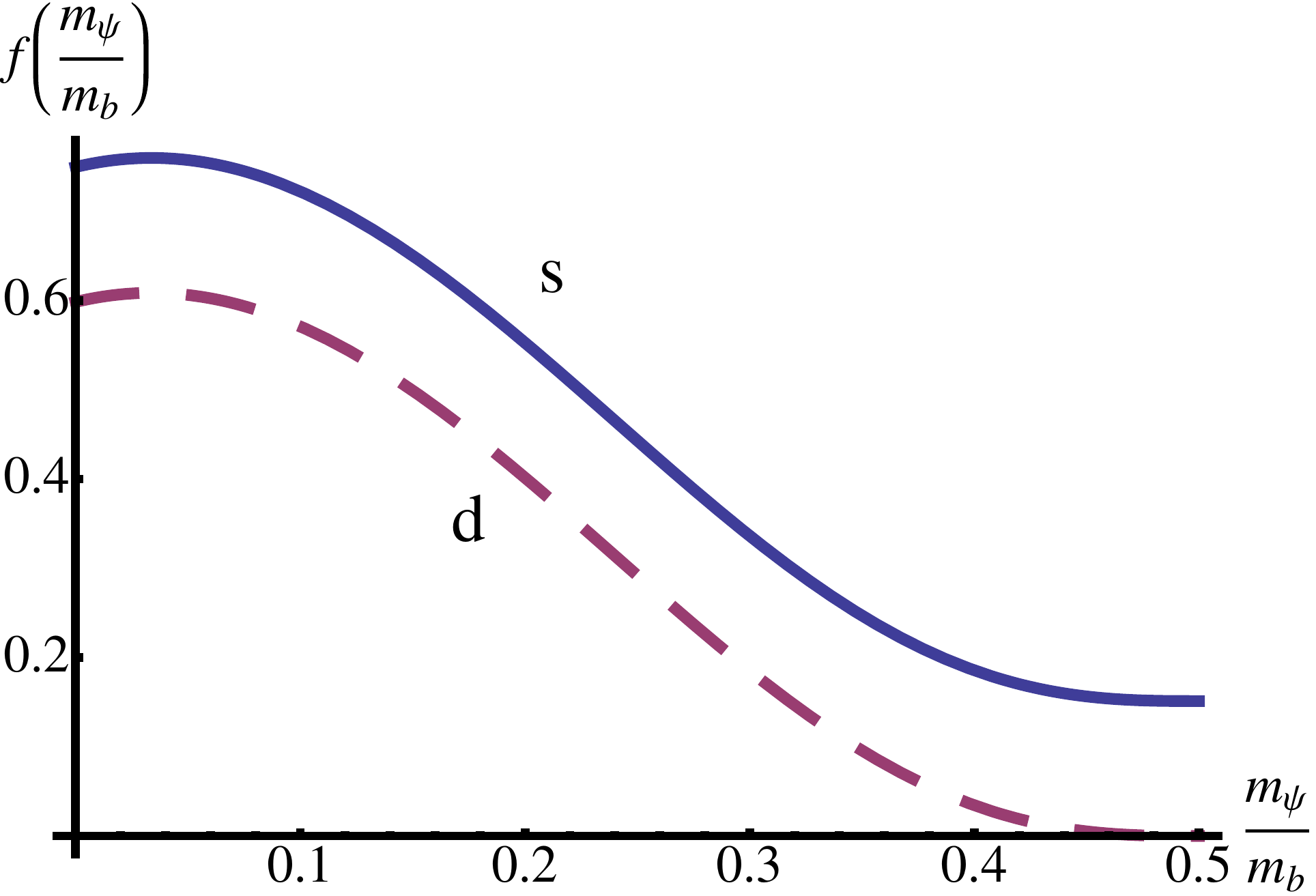}
\caption{The functions $f_{d,s}(m_\psi/m_b)$}
\label{fig:ffunct}
\end{figure}

It is important to note that  the operator $O^{s}_{\rm NP}$ will contribute to $B_d$ decays in addition to $B_s$ decays through the parton-level process $b\to s\bar\psi\psi$. While this decay is phase space suppressed, it is enhanced by two powers of $m_{B_s}/f_{B_s}$ and is in general the dominant contribution to the $B_s$ width from this new operator. We can estimate its effect on the $B_d$ and $B_s$ widths, including phase space factors, by writing
\begin{align}
&\frac{\Gamma_{\rm NP}^{d,s}}{\Gamma_{\rm tot}^{d,s}} \sim \frac{\Delta\Gamma_{\rm NP}^s}{\Delta\Gamma_{\rm tot}^s} \times f_{d,s}(m_\psi/m_b)\,.
\end{align}
Using the phase space factors given in~\cite{Bagan:1994qw}, an estimate of the function $f(m_\psi/m_b)$ is shown in Fig.~\ref{fig:ffunct}. The difference between the two functions is dominated by the annihilation contribution of $O^{s}_{\rm NP}$, which only affects the $B_s$ system. The exception is a new $\bar bs\bar dd$ operator, which contributes roughly equally to both $B_d$ and $B_s$ decay.

One can easily see that for light fields $\psi$ the contribution to the total lifetime of the $B_d$ and $B_s$ mesons can be as large as 50\% or more. Given that these lifetimes have been measured to 1\% and 2\% accuracy, respectively, one might conclude that NP contributions to $\Gamma_{12}^s$ are completely ruled out. However, our ability to predict these lifetimes accurately is plagued by large non-perturbative effects, and the resulting theoretical uncertainties could be as large as 10-20\%. Therefore, we will attempt to find alternate bounds on NP contributions to these operators, especially in the cases where $\psi = c \thinspace\text{or}\thinspace \tau$, for which the contributions are suppressed by phase space factors.

To constrain these operators, we need to consider how they contribute to observable decays of the $B$ mesons. Any non-leptonic operator of the form $\bar b q_1 \bar q_2 q_3$ will contribute to non-leptonic $B$ decay $B \to M_1 M_2$, where the flavors of $M_1$ and $M_2$ depend on the flavors of the quarks $q_i$. While non-perturbative effects make it difficult to predict the precise rate for non-leptonic decays, factorization theorems exist at leading order in $1/m_b$\cite{Beneke:2000ry,Bauer:2001cu}. This allows us to estimate the decay rates as
\begin{eqnarray}
{\rm Br}(B\to M_1 M_2) &\sim& \tau_B  G_F^2 |C|^2 \frac{f_M^2 m_b^3 F_{B \to M}}{32 \pi}\nn\\
&\sim& 10^{-3}\,,
\end{eqnarray}
where $C$ denotes the Wilson coefficient of the given operator. Here we have used the scaling $C \sim\lambda^2$ and the rough estimates $f_M \sim 0.15\thinspace {\rm GeV}$ and $F_{B \to M} \sim 0.3$ to obtain a numerical value for the decay rate.
Leptonic operators will contribute to decays of the form $B \to M \ell^+ \ell^-$, with a branching ratio estimated to be~\cite{Anikeev:2001rk}
\begin{eqnarray}
{\rm Br}(B\to M_1 \ell^+ \ell^-) &\sim& \tau_B  G_F^2 |C|^2 \, \frac{F_{B \to M} m_b^5}{192 \pi^3} \, {\rm PS}(m_\ell/m_b) \nn\\
& \sim& 0.02 \, {\rm PS}(m_\ell/m_b) \,,
\end{eqnarray}
One finds Êfor the phase space factor ${\rm PS}(0) = 1$ and Ê${\rm PS}(m_\tau/m_b) = 0.05$. They also contribute to the annihilation decay~\cite{Anikeev:2001rk}
\begin{eqnarray}
{\rm Br}(B\to \ell^+ \ell^-) &\sim& \tau_B  G_F^2 |C|^2 \frac{f_B^2 m_b^3}{32 \pi} H(m_\ell/m_b) \nn\\
&\sim&  0.3 \, H(m_\ell/m_b)\,,
\end{eqnarray}
where $H(m_\ell/m_b)$ is a helicity suppression factor that is $m_\ell^2/m_b^2$ if the the decay is helicity suppressed and unity otherwise, and we have used $f_B \sim 0.24 \, {\rm GeV}$ for the numerical estimate. 

Finally, operators of the form $\bar bs\bar\psi\psi$ can also contribute to the decay $B\to X_s\gamma$ by mixing into the operator $O_7$, which mediates this decay. If the NP operator is an (axial)-vector current, this mixing occurs at ${\cal O}(\alpha_s)\left({\cal O}(\alpha)\right)$, but is enhanced by large logarithms of $m_b/m_{t,W}$. Numerically, a non-leptonic (axial)-vector operator with a coefficient of order $G_F V_{cb}$ gives a contribution to $B\to X_s\gamma$ of the same order as the SM contribution, while a leptonic operator would be suppressed by a factor of $\alpha/\alpha_s$. If there is a (pseudo)scalar or tensor contribution, the mixing can occur at leading order and is still enhanced by large logarithms~\cite{Hiller:2003js}. This gives a contribution which is enhanced by $4\pi/\alpha_s$ compared to the SM contribution\footnote{We thank Uli Haisch for discussions on this point.}. Of course, one can make the helicity structure of the NP operator such that mixing into $O_7$ is forbidden, but such operators will mix instead into an $O_7'$, which still contributes to $B\to X_s\gamma$. The resulting contribution to the branching ratio can be estimated as
\begin{eqnarray}
\label{bsgammaconstraint}
\frac{{\rm Br}(B\!\!\to\!\! M \gamma)^{\rm NP}}{{\rm Br}(B\!\!\to\!\! M \gamma)^{\rm SM}} \!\sim\! 2 r \frac{C_{\rm NP}^{s}}{V_{cb}} +r^2\!\!\left(\frac{C_{\rm NP}^s}{V_{cb}}\right)^2 \!\!+ r^2\!\!\left(\frac{\widetilde C_{\rm NP}^{s}}{V_{cb}}\right)^2\!\!\!\!,
\end{eqnarray}
where  $C_{\rm NP}^{s}$ is the Wilson coefficient of the operator that mixes with the SM $O_7$ and $\widetilde C_{\rm NP}^{s}$ corresponds to the operator that mixes with $O'_7$. The variable $r$ is 1 for non-leptonic (axial)-vector operators, $\alpha/\alpha_s$ for leptonic (axial)-vector operators, and $4\pi/\alpha_s$ for (pseudo)scalar and tensor operators.

Given these results, one can immediately rule out the operators with $\psi=e,\mu$, since the observed branching ratios of $B \to K^{(*)} l^+ l^-$ are measured to be ${\cal O}(10^{-7})$, not ${\cal O}(10^{-2})$ as the presence of new operators would predict. The same is true for $\psi = \nu$, due to the limit on the branching ratio $B \to K^{(*)} \bar \nu \nu$. Thus, of the leptonic operators, only $\psi = \tau$ is allowed. 
The non-leptonic operators with light quarks are all ruled out by the absence of any 2-body non-leptonic $B$ decays to light mesons (such as $K$ and $\pi$) at the $10^{-3}$ level. This also rules out the combination of $\bar bs\bar cd$ and $\bar bs\bar dc$. The operator $\bar b s \bar c c$~\cite{Badin:2007bv}, however, cannot be excluded by this argument, since there is an SM contribution to this operator at the same level and the presence of non-perturbative effects makes a detailed comparison difficult. 

Both the remaining cases $(\psi = \tau,c)$ can be constrained by considering their contributions to the decay $B \to X_s\gamma$. Given that the measured value~\cite{Barberio:2008fa} of Br$^{\rm ex}(B\to X_s\gamma) = 3.52\pm0.25\times 10^{-4}$ is consistent with the theoretical prediction~\cite{Misiak:2006zs} of Br$^{\rm th}(B\to X_s\gamma) = 3.15\pm0.23\times10^{-4}$, only an ${\cal O}(10\%)$ correction can be accommodated. 
This eliminates any operator of the form $(\bar b s)(\bar \psi \psi)_{S,P,T}$, due to the factor $4 \pi / \alpha_s$ in Eq.~(\ref{bsgammaconstraint}). Note that this includes the operator discussed in~\cite{Dighe:2010nj}. 

The operators $(\bar b s)(\bar \tau \tau)_{V,A}$ can not be constrained because, as discussed, the mixing only occurs at one loop and is suppressed by $\alpha / \alpha_s$. The operator $(\bar b s)_{V-A}(\bar c c)_{V \pm A}$, which mixes with the operator $O_7$, can be eliminated, since its contribution to the decay $B \to X_s \gamma$ is of order $\left|\Gamma_{12}^{\rm NP}\right|/\left|\Gamma_{12}^{\rm SM}\right|$. An operator that mixes only with $O'_7$, on the other hand, contributes only quadratically to $B \to X_s \gamma$, and therefore $C_{\rm NP}^{s}/V_{cb} \sim 0.3$ would still be allowed. For this to lead to a sizable effect in $\Gamma_{12}^s$, the operator has to interfere with the SM operator $(\bar b c)_{V-A} (\bar cc)_{V-A}$ in its contribution to $\Gamma_{12}^s$. An operator with  helicity structure  ($\bar bs)_{V+A}(\bar cc)_{V-A}$ has this property and can therefore contribute significantly to the lifetime difference in the $B_s$ system.
%The constraints discussed here are true for vector and axial vector operators. As discussed above, the constraints from scalar, pseudoscalar, and tensor operators are even stronger than these. 

We have shown that there are only two possible SM operators that can give rise to an ${\cal O}(1)$ change in $\Gamma_{12}^s$. The first is  $(\bar bs)(\bar \tau\tau)_{V,A}$. This operator can be constrained by both $B\to K^{(*)}\tau^+\tau^-$ and $B_s\to\tau^+\tau^-$; however, due to the difficulty in detecting $\tau$'s, there is currently no bound on either decay. We therefore find that $\bar bs\bar\tau\tau$ can contribute significantly to to $\Gamma_{12}^s$. The second possible operator is of the form $(\bar bs)_{V+A}(\bar cc)_{V-A}$. 

Note that both of these operators would give rise to an order 10\% contribution to the total lifetime of the $B_s$ meson, if we require that they contribute an $O(1)$ amount to $\Gamma_{12}^s$. As discussed above, however, this does not contradict the precise measurement of the $B_s$ lifetime, due to the large theoretical uncertainties when trying to predict this quantity. On the other hand, the ratio $\tau_{B_s}/\tau_{B_d}$ is under much better theoretical control. This is because the unknown nonperturbative effects largely cancel in the ratio, such that it can be predicted with high accuracy~\cite{Neubert:1996we}
\begin{equation}
\frac{\tau(B_s)}{\tau(B_d)} = 1 \pm O(1\%)\,.
\end{equation}
An operator that gives an $\mathcal{O}(1)$ contribution to $\Gamma_{12}^s$ would give rise to large lifetime difference $1- \tau(B_s)/\tau(B_d) = O(10\%)$, much larger than the theoretical uncertainty in this quantity. Unfortunately, the experimental uncertainties in the ratio~\cite{HFAG}
\begin{equation}
\frac{\tau(B_s)}{\tau(B_d)} = 0.965 \pm 0.017
\end{equation}
are somewhat larger than our theoretical knowledge. While one can rule out a 10\% effect, a 5\% contribution is still allowed. In fact, the current measurement seems to indicate a 2$\sigma$ deviation in the lifetime ratio. If a significant difference from unity of this lifetime ratio could be established, it would be another hint at new physics contributing to  $\Gamma_{12}^s$. A more precise measurement of this quantity is therefore of great importance.

What would one conclude if a new measurement of this lifetime ratio does not allow for a large deviation from unity?
Since the operators discussed above reduce the ratio $B_s$ lifetime relative to the $B_d$ lifetime, one would be forced to add new operators of the form
\begin{equation}
O^{d}_{\rm NP}=\bar b\Gamma d\bar\psi_1\Gamma\psi_2
\end{equation}
with Wilson coefficient
\begin{equation}
C_{\rm NP}^{d} \simeq C_{\rm NP}^{s}\,,
\end{equation}
to make up for this difference. In general, we do not need the fields in $O^{d}_{\rm NP}$ to be the same as in $O^{s}_{\rm NP}$, which allows $\psi \neq \psi_1 \neq \psi_2$. Note that as long as $\psi_1 \neq \psi_2$, such operators would not contribute to $\Gamma_{12}^d$.  While one could potentially add several new operators to the $B_d$ sector with smaller Wilson coefficients to compensate for $O^{s}_{\rm NP}$, we will assume that only one new operator is added. Our arguments can be easily generalized to the case of multiple operators. Adding operators to the $B_d$ system would also change the ratio $\tau\left(B_u\right)/\tau\left(B_d\right)$, which is also well understood theoretically \cite{Beneke:2002rj},~\cite{Franco:2002fc}. However, we will not explore new operators in the $B_u$ system here.

The possible operators are shown on the right hand side of Table~\ref{tab:operators}. As before, the operators $\bar b d \bar ll$ for $l=e,\mu$ are ruled out by limits on the decay $B \to \pi l^+ l^-$, and the operator with $\psi = \nu$ by the limit on $B \to \pi \bar \nu \nu$. The operator $\bar b d \bar \tau\tau$ is in this case excluded by the experimental limit on the decay $B\to \tau^+ \tau^-$. Operators involving only light quarks are also ruled out, again due to the absence of 2-body non-leptonic $B$ decays to light mesons at the $10^{-3}$ level. The operator $\bar b d \bar cu$ (which has the wrong sign charm) is also excluded by this argument, while the operator $\bar bd\bar cc$ is ruled out by its large contribution to $B\to(\rho,\omega^0)\gamma$. 

This leaves only one operator one could add to the $B_d$ system, namely $\bar bd\bar uc$. A new physics operator with $C^{d}_{\rm NP}\!\sim\!\lambda^2$ would contribute at the same order as the SM operator. While this would increase the rate of non-leptonic $B$ decays of the form $B \to D \pi$, the presence of non-perturbative effects makes it difficult to rule out this operator conclusively. Note that \ this operator makes it possible to keep the lifetime ratio between $B_s$ and $B_d$ mesons the same, as long as its coefficient is tuned sufficiently. Given that the operator $\bar bd\bar uc$ appears to be completely unrelated to either the $\bar bs \bar \tau\tau$ or $(\bar bs)_{V+A}(\bar cc)_{V-A}$ operators allowed in the $B_s$ system, such a NP scenario seems very contrived.

Having discussed in detail the effect of dimension six operators containing SM fields, we want to briefly comment on other possibilities for NP, while stressing that we can not rigorously discuss all possible extensions. The simplest extension would be to allow for new light degrees  of freedom (denoted by $\phi$ and $\psi$ for spin 0 and spin 1/2), that are SM singlets (or electrically neutral, but charged under SU(2)$_{\rm L}$). Operators of this form include $\bar b(s,d)\phi\phi$ and $\bar b(s,d)\bar\psi\psi$,  $\bar b(s,d)\phi F$ and  $\bar b(s,d) \phi G$ where $F$ and $G$ denote the photon and gluon field strength, respectively. The first two operators are ruled out by the experimental bounds on $B\to(K^{(*)},\omega,\rho)\bar\nu\nu$. The operator $\bar bd\phi F$ is ruled out by the absence of the decay $B \to \bar\nu\nu \gamma$, but the other operators are not ruled out, at least not without a more careful analysis. 

As another possibility, one might consider higher dimensional operators containing SM fields. At dimension seven, a derivative could be added to any of the operators we have already discussed. All of the previously mentioned constraints would still need to be considered; in addition, it is unlikely that dimension seven operators could contribute at the same level as the dimension six operators. 
However, three new operators appear: $\bar b(s,d)FF$, $\bar b(s,d)GG$ and $\bar b(s,d)FG$, where $F$ and $G$ are the photon and gluon field strengths respectively. The operators with two photons are ruled out by the decays $B_{d,s}\to\gamma\gamma$ and $B\to X_{d,s}\gamma$. The operators with two gluons would generate four quark operators of the form $\bar b(s,d)\bar qq$ of roughly the same size as the previously studied dimension six operators and are thus excluded. Finally, the operators with one photon and one gluon also contribute to $B\to X_{d,s}\gamma$, and are therefore ruled out. 
One could continue to add higher dimensional SM operators, but given that we need a fairly large contribution to $\Gamma_{12}^s$, it is unlikely that this could be done without lowering $\Lambda_{\rm NP}$ into a region that is excluded by collider experiments. 

As a last possibility, we would like to comment on operators  that contain one or more new light degrees of freedom, but which eventually will decay to SM particles. Assuming that light degrees of freedom that are charged under SU(3)$_{\rm C}$ or have ${\cal O}(e)$ electric charge are already ruled out by collider data, we are left with fields that couple to $\bar b(s,d)$ through higher dimensional operators. They  also have to couple to the Standard Model in such a way that they predominantly decay to particles charged under the SM before exiting the detector, without having any ${\cal O}(1)$ SM charges. Finally, these operators are constrained to contribute to $\Gamma_{s,d}$ at no more than the 20\% level and maintain the measured ratio $\tau_{B_s}/\tau_{B_d}$. It might be an interesting exercise to see if such a scenario can be realized by clever model building. 

In conclusion, we have analyzed if current data allows significant new physics contributions to $\Gamma_{12}^s$, as recent measurements might suggest. We have shown that there are two dimension six operators in the Standard Model which are still allowed given current experimental constraints. More detailed measurements of the decays $B \to X_s \tau^+ \tau^-$ and $B_s \to \tau^+ \tau^-$ would establish the presence of one of these operators or rule it out. Both of these operators would contribute differently to the $B_s$ and the $B_d$ lifetime. This ratio is predicted very accurately in the SM, and an observed value in agreement with the Standard Model prediction (and similarly small uncertainties) would require additional physics in the $B_d$ system in a way that seems unrelated to the new physics allowed in the $B_s$ system. Therefore, a more precise measurement of $\tau(B_s)/\tau(B_d)$, which is possible with currently available data, could either support or strongly constrain the existence of new physics in $\Gamma_{12}^s$.

%%%%%%%%%%%%%%%%%%%%%%%%%%%%%%%%%%%%%%%%%%%%%%%%%%%%%%%%%%%%%%%%%%%%%%%%%%%%%%%%
\begin{acknowledgments}
This work was supported by the Director, Office of Science, Offices of High Energy and Nuclear Physics of the U.S. 
Department of Energy under the Contracts DE-AC02- 05CH11231. We would like to thanks Asimina Arvanitaki, Clifford Cheung and Piyush Kumar for collaboration at early stages of this work, and Zoltan Ligeti for many stimulating discussions. CWB would like to acknowledge support from the Aspen Center for Physics, where much of this work was performed. 
\end{acknowledgments}

%%%%%%%%%%%%%%%%%%%%%%%%%%%%%%%%%%%%%%%%%%%%%%%%%%%%%%%%%%%%%%%%%%%%%%%%%%%%%%%%


\begin{thebibliography}{10}
\bibitem{CDF9015} CDF Collaboration, ``Measurement of CP asymmetry in semileptonic $B$ decays'', Note 9015, Oct. 2007

\bibitem{Abazov:2010hv}
  V.~M.~Abazov {\it et al.}  [The D0 Collaboration],
  %``Evidence for an anomalous like-sign dimuon charge asymmetry,''
  arXiv:1005.2757 [hep-ex].
  %%CITATION = ARXIV:1005.2757;%%

\bibitem{Abazov:2009wg}
  V.~M.~Abazov {\it et al.}  [D0 Collaboration],
  %``Search for CP violation in semileptonic Bs decays,''
  arXiv:0904.3907 [hep-ex].
  %%CITATION = ARXIV:0904.3907;%%

\bibitem{Barberio:2008fa}
  E.~Barberio {\it et al.}  [Heavy Flavor Averaging Group],
  %``Averages of $b-$hadron and $c-$hadron Properties at the End of 2007,''
  arXiv:0808.1297 [hep-ex].
  %%CITATION = ARXIV:0808.1297;%%

\bibitem{Lenz:2006hd}
  A.~Lenz and U.~Nierste,
  %``Theoretical update of $B_s - \bar{B}_s$ mixing,''
  JHEP {\bf 0706}, 072 (2007)
  [arXiv:hep-ph/0612167].
  %%CITATION = JHEPA,0706,072;%%

\bibitem{latticeaverages} For a collection of recent lattice results, see {\tt www.latticeaverages.org}

\bibitem{Grinstein:2001zq}
For a discussion of heavy flavor decays in the t'Hooft model, see
  B.~Grinstein,
  %``Global duality in heavy flavor decays in the 't Hooft model,''
  Phys.\ Rev.\  D {\bf 64}, 094004 (2001)
  [arXiv:hep-ph/0106205];
  Phys.\ Lett.\  B {\bf 529}, 99 (2002)
  [arXiv:hep-ph/0112323].
  %%CITATION = PHRVA,D64,094004;%%

\bibitem{Grossman:2009mn}
  Y.~Grossman, Y.~Nir and G.~Perez,
  %``Testing New Indirect CP Violation,''
  Phys.\ Rev.\ Lett.\  {\bf 103}, 071602 (2009)
  [arXiv:0904.0305 [hep-ph]].
  %%CITATION = PRLTA,103,071602;%%

\bibitem{CDFD0combination} see {\tt http://www-cdf.fnal.gov/physics/new/\\
bottom/090721.blessed-betas\_combination2.8/\\D0Note5928\_CDFNote9787.pdf}

\bibitem{HFAG}
{\tt http://www.slac.stanford.edu/xorg/hfag/}

\bibitem{Ligeti:2010ia}
  Z.~Ligeti, M.~Papucci, G.~Perez and J.~Zupan,
  %``Implications of the dimuon CP asymmetry in B_{d,s} decays,''
  arXiv:1006.0432 [hep-ph].
  %%CITATION = ARXIV:1006.0432;%%

\bibitem{Dobrescu:2010rh}
  B.~A.~Dobrescu, P.~J.~Fox and A.~Martin,
  %``CP violation in B_s mixing from heavy Higgs exchange,''
  arXiv:1005.4238 [hep-ph].
  %%CITATION = ARXIV:1005.4238;%%

\bibitem{Dighe:2007gt}
 A.~Dighe, A.~Kundu and S.~Nandi,
 %``Possibility of large lifetime differences in neutral $B$ meson systems,''
 Phys.\ Rev.\  D {\bf 76}, 054005 (2007)
 [arXiv:0705.4547 [hep-ph]].
 %%CITATION = PHRVA,D76,054005;%%

\bibitem{Dighe:2010nj}
  A.~Dighe, A.~Kundu and S.~Nandi,
  %``Enhanced $\bsbsbar$ lifetime difference and anomalous like-sign dimuon
  %charge asymmetry from new physics in $B_s \to \tau^+ \tau^-$,''
  arXiv:1005.4051 [hep-ph].
  %%CITATION = ARXIV:1005.4051;%%
  
\bibitem{Randall:1998te}
 L.~Randall and S.~f.~Su,
 %``CP violating lepton asymmetries from $B$ decays and their implication for
 %supersymmetric flavor models,''
 Nucl.\ Phys.\  B {\bf 540}, 37 (1999)
 [arXiv:hep-ph/9807377].
 %%CITATION = NUPHA,B540,37;%%

\bibitem{Buras:2010mh}
  A.~J.~Buras, M.~V.~Carlucci, S.~Gori and G.~Isidori,
  %``Higgs-mediated FCNCs: Natural Flavour Conservation vs. Minimal Flavour
  %Violation,''
  arXiv:1005.5310 [hep-ph].
  %%CITATION = ARXIV:1005.5310;%%

\bibitem{Jung:2010ik}
  M.~Jung, A.~Pich and P.~Tuzon,
  %``Charged-Higgs phenomenology in the Aligned two-Higgs-doublet model,''
  arXiv:1006.0470 [hep-ph].
  %%CITATION = ARXIV:1006.0470;%%

\bibitem{Bagan:1994qw}
  E.~Bagan, P.~Ball, V.~M.~Braun and P.~Gosdzinsky,
  %``Theoretical update of the semileptonic branching ratio of B mesons,''
  Phys.\ Lett.\  B {\bf 342}, 362 (1995)
  [Erratum-ibid.\  B {\bf 374}, 363 (1996)]
  [arXiv:hep-ph/9409440].
  %%CITATION = PHLTA,B342,362;%%

\bibitem{Beneke:2000ry}
  M.~Beneke, G.~Buchalla, M.~Neubert and C.~T.~Sachrajda,
  %``QCD factorization for exclusive, non-leptonic B meson decays: General
  %arguments and the case of heavy-light final states,''
  Nucl.\ Phys.\  B {\bf 591}, 313 (2000)
  [arXiv:hep-ph/0006124].
  %%CITATION = NUPHA,B591,313;%%

\bibitem{Bauer:2001cu}
  C.~W.~Bauer, D.~Pirjol and I.~W.~Stewart,
  %``A proof of factorization for B --> D pi,''
  Phys.\ Rev.\ Lett.\  {\bf 87}, 201806 (2001)
  [arXiv:hep-ph/0107002];
  %%CITATION = PRLTA,87,201806;%%
    C.~W.~Bauer, D.~Pirjol, I.~Z.~Rothstein and I.~W.~Stewart,
  %``B --> M(1) M(2): Factorization, charming penguins, strong phases, and
  %polarization,''
  Phys.\ Rev.\  D {\bf 70}, 054015 (2004)
  [arXiv:hep-ph/0401188].
  %%CITATION = PHRVA,D70,054015;%%

 \bibitem{Anikeev:2001rk}
  K.~Anikeev {\it et al.},
  %``$B$ physics at the Tevatron: Run II and beyond,''
  arXiv:hep-ph/0201071.
  %%CITATION = HEP-PH/0201071;%%

\bibitem{Hiller:2003js}
  G.~Hiller and F.~Kruger,
  %``More model independent analysis of $b \to s$ processes,''
  Phys.\ Rev.\  D {\bf 69}, 074020 (2004)
  [arXiv:hep-ph/0310219].
  %%CITATION = PHRVA,D69,074020;%%

\bibitem{Badin:2007bv}
  A.~Badin, F.~Gabbiani and A.~A.~Petrov,
  %``Lifetime difference in $B_s$ mixing: Standard model and beyond,''
  Phys.\ Lett.\  B {\bf 653}, 230 (2007)
  [arXiv:0707.0294 [hep-ph]].
  %%CITATION = PHLTA,B653,230;%%

 \bibitem{Misiak:2006zs}
  M.~Misiak {\it et al.},
  %``The first estimate of B(anti-B --> X/s gamma) at O(alpha(s)**2),''
  Phys.\ Rev.\ Lett.\  {\bf 98}, 022002 (2007)
  [arXiv:hep-ph/0609232].
  %%CITATION = PRLTA,98,022002;%%

\bibitem{Neubert:1996we}
  M.~Neubert and C.~T.~Sachrajda,
  %``Spectator effects in inclusive decays of beauty hadrons,''
  Nucl.\ Phys.\  B {\bf 483}, 339 (1997)
  [arXiv:hep-ph/9603202].
  %%CITATION = NUPHA,B483,339;%%

\bibitem{Esen:2010jq}
  S.~Esen {\it et al.},
  %``Observation of Bs->Ds(*)+Ds(*)- using e+e- collisions and a determination
  %of the Bs-Bsbar width difference \Delta\Gamma_s,''
  arXiv:1005.5177 [hep-ex].
  %%CITATION = ARXIV:1005.5177;%%
  
\bibitem{Beneke:2002rj}
  M.~Beneke, G.~Buchalla, C.~Greub, A.~Lenz and U.~Nierste,
  %``The B+ - B/d0 lifetime difference beyond leading logarithms,''
  Nucl.\ Phys.\  B {\bf 639}, 389 (2002)
  [arXiv:hep-ph/0202106].
  %%CITATION = NUPHA,B639,389;%%
  
\bibitem{Franco:2002fc}
  E.~Franco, V.~Lubicz, F.~Mescia and C.~Tarantino,
  %``Lifetime ratios of beauty hadrons at the next-to-leading order in QCD,''
  Nucl.\ Phys.\  B {\bf 633}, 212 (2002)
  [arXiv:hep-ph/0203089].
  %%CITATION = NUPHA,B633,212;%%


\end{thebibliography}
\end{document}